\documentclass[12pt]{article}

\usepackage{newtxtext,newtxmath}

\usepackage{amsmath, amssymb, bbold}
\usepackage{mathtools}
\usepackage[arrowdel]{physics}
\usepackage{graphicx,color,psfrag}
\usepackage{ulem}
\usepackage{tikz}
\usepackage{hyperref}

\usepackage{graphicx}

\usepackage[letterpaper,margin=1in]{geometry}

\renewenvironment{abstract}
	{\quotation}
	{\endquotation}

\date{}

\makeatletter
\renewcommand{\fnum@figure}{\textbf{Figure \thefigure}}
\renewcommand{\fnum@table}{\textbf{Table \thetable}}
\makeatother

\usepackage{scicite}

\usepackage{url}

\newcommand{\bx}{\mathbf x}

\newcommand{\bA}{\mathbf A}

\newcommand{\ci}{\mathfrak i}

\newcommand{\bj}{\mathbf j}

\newcommand{\kB}{k_\text{B}}

\newcommand{\mfh}{\mathfrak{h}}

\newcommand{\cev}[1]{\reflectbox{\ensuremath{\vec{\reflectbox{\ensuremath{#1}}}}}}

\def\scititle{
Entropy Flow at the Quantum Limit
}
\title{\bfseries \boldmath \scititle}

\author{
    Marco A.\ Jimenez-Valencia$^{1}$,
    Parth Kumar$^{1\ast}$,
    Yiheng Xu$^{2}$,\and
    Ferdinand Evers$^{3}$,
    Charles A.\ Stafford$^{1}$\and    
    \small$^{1}$Department of Physics, University of Arizona, Tucson, Arizona 85721, USA.\and
	\small$^{2}$Department of Physics, University of California, San Diego, California 92093, USA.\and
    \small$^{3}$Institute of Theoretical Physics, University of Regensburg, D-93050 Regensburg, Germany.\and
	\small$^\ast$Corresponding author. Email: parthk@arizona.edu%
}

\begin{document} 

\maketitle

\begin{abstract} \bfseries \boldmath
Thermal management is a key challenge, both globally and microscopically in integrated circuits and quantum technologies.
The associated heat flow $I_Q$ has been understood 
since the advent of thermodynamics by a process of elimination, $I_Q{=}I_E{-}\mu I_N$, subtracting from the energy flow $I_E$ its convective contribution. %
However, in the quantum limit, this formula implies the paradoxical result that the entropy entrained by heat flow is unbounded even though the entropy itself tends to zero.
We resolve this conundrum by recognizing that the traditional formula for heat is missing a quantum term.
The correct quantum formula predicts that the heat produced in quantum processes is vastly smaller than previously believed, with correspondingly beneficial consequences for the efficiency of quantum machines.
\end{abstract}

\noindent
The dissipation of heat places fundamental limits on all physical processes, including the operation of quantum machines such as quantum computers.  Without an understanding of heat and entropy flow at the quantum limit, it is impossible even to estimate the performance of such quantum machines, if they ever were to be practical.
Although heat in macroscopic systems has been understood since Robert Mayer in the 19th century and the entropy of quantum systems was formulated in  1932 by John von Neumann \cite{vonneumannMathematicalFoundationsQuantum2018}, the flow of entropy and heat at the quantum level has never been properly formulated theoretically, and its experimental exploration is still in its infancy \cite{Lee2013,mengesTemperatureMappingOperating2016a,Reddy2017,Menges2017,kleeorinHowMeasureEntropy2019,childEntropyMeasurementStrongly2022}.

The textbook expression \cite{mahanManyParticlePhysics2000}  for heat flow $I_Q$ equates it to the difference between the energy flow $I_E$ and its convective component, also known as the (rate of) chemical work 
$
    I_Q=I_E-\mu I_N,
    $
where $\mu$ is the (electro)chemical potential of the system and $I_N$ is the flow of particles.  Quite generally for quasi-reversible processes, the flow of heat is related to the flow of entropy by $I_Q=T I_S$, where $T$ is the temperature of the system.  However, as we demonstrate below, the textbook formula 
leads to the bizarre %
conclusion that $I_S \rightarrow \pm \infty$ as $T\rightarrow 0$, in flagrant violation of the 3rd Law of Thermodynamics. This paradoxical behaviour originates from a missing term %
that must be further subtracted from the energy flow %
in order to give a heat flow consonant with the Laws of Thermodynamics.

In this research article, we present a consistent quantum theory of heat flow that allows us to derive the heretofore missing term.  Importantly, our theory predicts that the heat dissipated in quantum processes at low absolute temperatures is orders of magnitude less than that predicted by the conventional formula, with highly beneficial consequences for the potential feasibility of quantum technology.

\section*{The flow of heat, work, and entropy %
}
Let us recapitulate: In textbook thermodynamics the heat flow  is understood as one out of three components that are associated with the flow of energy. Namely, when a system like a heat engine loses an amount of energy per time, $\dot E$, this energy can be given away in the form of chemical work, $\dot W_N$, or of external work, 
$\dot W$, e.g. driving a wheel of a steam engine; the remainder of the energy loss constitutes the heat transferred to the environment: $\dot Q = \dot E - \mu\dot N  - \dot W$. We here cite the relation in a form that pays tribute to the fact that the chemical work is closely related to the change of the system's particle number per time: $\dot W_N=\mu \dot N$. 

One might expect - and  we will here show it is indeed the case - that the fundamental formula for the loss rates carries over in a one-to-one manner  to the associated flows of heat, energy, particle number and work:
$I_Q = I_E - \mu I_N - I_\Omega$. However 
in the conventional expression for heat flow, the third term, $I_\Omega$, is not foreseen. It represents the flow of useful work, which can also be seen as a flow of free energy. 
\begin{figure}[b]
    \centering
    \includegraphics[width=0.9\linewidth]{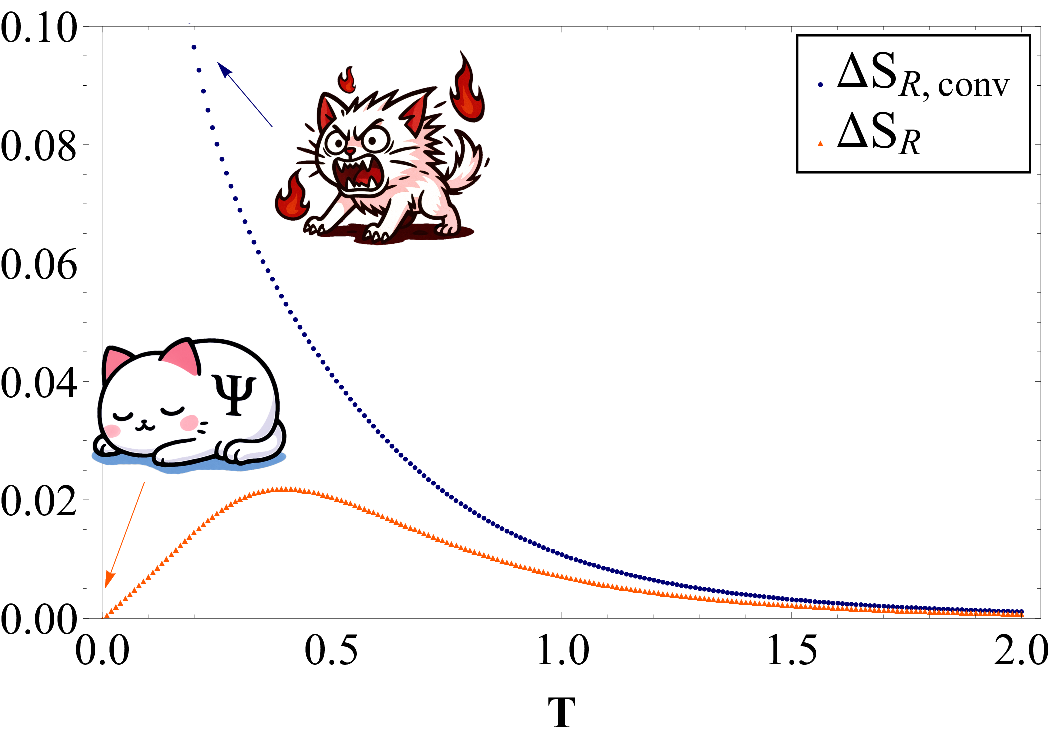}
    \caption{{\bf Comparison of the correct and conventional entropy formulae.}  The entropy $\Delta S_R$ (red triangles) produced in the reservoir for a quasi-static process in the resonant level model, as a function of the temperature $T$ (conventional formula $\Delta S_{R,{\rm conv}}$ shown in blue dots). Here the level is driven $\varepsilon_s(t)$: $1 \rightarrow 1.5$, the  chemical potential of the reservoir is $\mu=0$ and the system-reservoir coupling is $V=1$. The hopping matrix element in the reservoir is $t_0=1.25$. The correct result is in accord with the 3rd Law of Thermodynamics, which holds that the entropy of the reservoir should tend to zero as $T\rightarrow 0$ (represented by a cat in its quantum mechanical ground state), while the conventional result implies a spurious divergent entropy as $T\rightarrow 0$ (represented by the cat in a high-entropy state).
    } %
    \label{fig_entropychange_comp}
\end{figure}

\begin{figure}
 \centering
\includegraphics[width=1\linewidth]{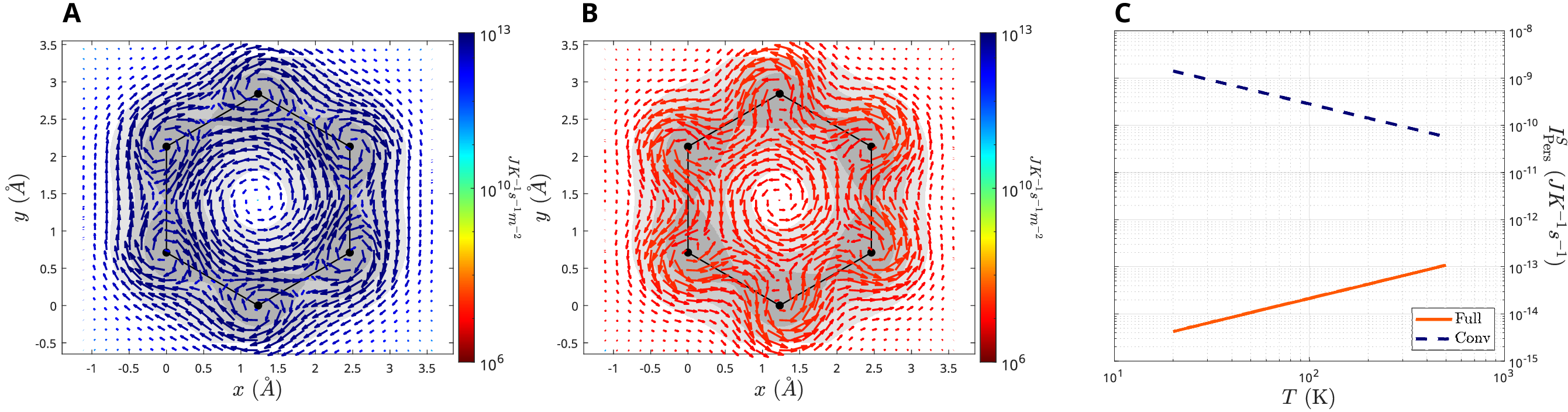}
\caption{{\bf Equilibrium entropy currents induced by a magnetic field in a benzene molecule.}
(\textbf{A}) Conventional entropy current density $\bj_{S,{\rm conv}}=(\bj_e-\mu \bj_n)/T$ flowing in a plane $0.5$\AA\ above the plane of the carbon nuclei in a benzene ring adsorbed on a metal surface and threaded by a magnetic flux $\phi=0.05 \phi_0$, where $\phi_0$ is the magnetic flux quantum. The $\pi$-electron density of the molecule is shown in gray scale.
(\textbf{B}) Correct quantum entropy current density $\bj_S=(\bj_e-\mu \bj_n-\bj_\Omega)/T$. (\textbf{C}) Total persistent entropy current comparison. Note that the conventional entropy current diverges at low temperatures in violation of the 3rd Law of Thermodynamics. The calculation was carried out in the H\"uckel model and in all plots except (C) $T= %
300K$. }
\label{f1}
\end{figure}

Before explaining why $I_\Omega$ has been overlooked since the invention of quantum mechanics a century ago, let us illustrate its significance with two striking examples %
in Figs.\ \ref{fig_entropychange_comp} and \ref{f1}. 
Fig.\ \ref{fig_entropychange_comp} shows the entropy produced in a reservoir when a resonant level coupled to it is driven quasi-statically. (For details of the model and calculation method, see Sec.\  \ref{app:nonlocal_work} in \cite{methods}.) The correct result $\Delta S_R=(\Delta E_R -\mu\Delta N_R -\Delta \Omega_R)/T$ is plotted in red, while the conventional formula $\Delta S_{R, {\rm conv}}=(\Delta E_R -\mu\Delta N_R)/T$ is plotted in blue.  The neglect of the free energy flow term, identified as {\it nonlocal quantum work} in Ref.\ \cite{kumarWorkSumRule2024a}, leads to a clear violation of the 3rd Law of Thermodynamics in the conventional formula.  %
The vastly smaller entropy production predicted by the correct quantum formula implies correspondingly higher efficiency bounds on such a quantum machine.
Related challenges in defining heat and entropy consistent with the 3rd Law for a variety of quantum processes are discussed in 
Refs.\   
\cite{nieuwenhuizenStatisticalThermodynamicsQuantum2002,fordEntropyQuantumOscillator2005,
oconnellDoesThirdLaw2006,levyQuantumRefrigeratorsThird2012,kolarQuantumBathRefrigeration2012,
cleurenCoolingHeatingRefrigeration2012,levyCommentCoolingHeating2012,kosloffQuantumThermodynamicsDynamical2013,ludovicoDynamicalEnergyTransfer2014,espositoNatureHeatStrongly2015,bruchQuantumThermodynamicsDriven2016,
masanesGeneralDerivationQuantification2017,
bruchLandauerButtikerApproachStrongly2018,
shastryThirdLawThermodynamics2019}.

Fig.\ \ref{f1} shows the equilibrium entropy current circulating in a ring-shaped quantum system (persistent current) in the presence of a magnetic field. 
(For details of the model and calculations, see Sec.\ \ref{app:SGF} in \cite{methods}.) Such currents can be driven around a benzene ring or larger annulene molecules \cite{Calder66,Schroeder66}, as indicated in Fig.\ \ref{f1}, but they can also flow in much larger mesoscopic systems \cite{Eckern2002}.  
Panel  (A) shows the conventional entropy current density $\bj_{s,{\rm conv}}=(\bj_e-\mu \bj_n)/T$,
which is the local variant of the textbook formula $T I_S = I_E - \mu I_N$; panel (B) shows the correct entropy current density
$\bj_s=(\bj_e-\mu \bj_n-\bj_\Omega)/T$.  
Comparing both figures, one observes that the current patterns deviate significantly. The most important deviation is, however, in the current magnitudes. 
To highlight this, we compare in panel (C) the conventional and correct formul\ae\ for the total entropy current circulating around the ring. The key message is that the amplitude of the conventional entropy current is several orders of magnitude larger than that of the correct quantum result and diverges in the limit $T\rightarrow 0$, while the correct quantum formula yields an entropy flow that vanishes in the limit $T\rightarrow 0$, consistent with the 3rd Law of Thermodynamics.
Figs.\ \ref{fig_entropychange_comp} and \ref{f1} demonstrate that the conventional formul\ae\ for heat and entropy flow are untenable at the quantum limit.

\section*{Entropy density and currents}

To answer the question of how the term in $I_\Omega$ could have been overlooked in the entropy current, we need to recall how in quantum mechanics the operator representations of current densities are usually derived. Three ingredients are needed: 
the definition of the density of a local observable, e.g. the particle density $\hat n(\bx)$, here given in operator form indicated by the hat accent; an equation of motion given in quantum mechanics by the Heisenberg %
equation, $-\ci \partial_t \hat n(\bx) = [\hat H, \hat n(\bx)]$; a continuity equation that relates the rate of density change to the divergence of a current density: $\partial_t \hat n(\bx)+\text{div} \hat \bj_n(\bx)=0$. While the procedure is successful for deriving particle ($\hat \bj_n(\bx)$) and energy 
($\hat \bj_e(\bx)$) current densities, it does have its intricacies. 
First, it is assumed that the conserved quantity is associated with a well defined local density operator. In local quantum field theories the corresponding densities are straightforward to define for particle and energy density not so, however, for the entropy density.  
Second, the continuity equation only fixes the vector fields $\hat \bj(\bx)$ up to a curl; ring currents require extra considerations. For instance, the definition of the charge current $e \hat \bj_n(\bx)$ is completed with respect to the curl only by invoking a gauge argument. We claim that the flow of useful work, $I_\Omega$, corresponds to a curl, which is missed in the textbook derivations.

To highlight why $I_\Omega$ is absent in the conventional expression for the entropy current in better detail, we recall Boltzmann's insight that the thermodynamic entropy is a statistical concept. When translating Boltzmann's result to quantum mechanics, the entropy is promoted to an operator: $\hat S = - \kB \ln \hat \rho$, with $\kB$ the Boltzmann constant; it is the logarithm of the density operator $\hat \rho$, which defines the weights of quantum states in the statistical ensembles. The ensemble of relevance for the 
equilibrium 
current flows 
shown in Fig.\ \ref{f1} 
is the grand-canonical one, which by definition has $\hat \rho = \exp{-(\hat H - \mu \hat N - \Omega)/\kB T}$; the operator $\hat H$ is the Hamilton operator and $\hat N$ the particle number operator of the quantum system. The entropy operator of this ensemble, being simply given by the logarithm of $\hat \rho$, is therefore given as $\hat S = (\hat H - \mu \hat N-\Omega)/T$, where $\Omega$ denotes the grand %
potential. 

\subsection*{Derivation of the entropy current density} 

From this starting point, the definition of the entropy density operator, $\hat s(\bx)$, is straightforward, because $\hat H, \hat N$, and also the free energy $\Omega$ are readily expressed as volume integrals over a local density, e.g. 
$\Omega = \int_{\mathcal{V}}\ d\bx\ \omega(\bx)$.
The resulting expression for $\hat s(\bx)$ is formulated most transparently using field operators, $\hat \psi^\dagger(\bx)$ and $\hat \psi(\bx)$; for free particles it reads  
\begin{align}
\label{eq:entropy_density_conv}
    T \hat s(\bx) \coloneqq \hat \psi^\dagger(\bx) (\mfh -\mu) \hat\psi(\bx) - \omega(\bx),  
\end{align}
where the first term represents the energy density, $\mfh$ denoting the single-particle Hamiltonian, and the second term represents $\mu\hat n(\bx)$. 
Embarking on this definition for $\hat s(\bx)$, the standard procedure for deriving current densities from divergences yields for the entropy current density the familiar relation 
$T \hat\bj_{s,{\rm conv}}(\bx) = \hat\bj_e(\bx) - \mu \hat\bj_n(\bx)$. 
A term representing $I_\Omega$ does not arise, because the commutator $[\hat H,\omega(\bx)]$ vanishes.

As is clear from the outset, the standard procedure identifies $\hat \bj_s(\bx)$ only up to a curl. Hence after replacing $\hat s(\bx)$ with a mathematically equivalent representation of 
the entropy density, %
the standard procedure can give a current density, $\hat \bj_s({\bx})$, that deviates from $\hat \bj_{s,{\rm conv}}(\bx)$ by a curl. Such an equivalent representation of the entropy density is readily found for free particles; for instance, for fermions it is given by 
\begin{align}
    \hat s(\bx) &\coloneqq - 
    k_B\left( \hat\psi^\dagger(\bx)\,\overset\longleftrightarrow{\ln f}\,\hat \psi(\bx)  +\hat \psi(\bx)\,\overset\longleftrightarrow{\ln  p}\,\hat \psi^\dagger(\bx)
    \right); 
    \label{e2} 
\end{align}
here $f$ and $p{=}1{-}f$ denote general distributions of particles and holes and 
$\overset\longleftrightarrow{Q}:=(\overset\longrightarrow{Q}+\overset\longleftarrow{Q})/2$, where the arrows indicate if the operator acts to the right or left.
In equilibrium, $f$ becomes the Fermi-Dirac distribution, $f({\mfh})$, and Eq.\  
\eqref{e2} reduces to Eq.\ \eqref{eq:entropy_density_conv}.

Embarking with \eqref{e2} on the standard procedure, one extracts an entropy current density 
that includes an extra term \cite{jimenez-valenciaPersistentDissipativePeltier2025a}, explicitly 
$
    \bj_\Omega(\bx) \coloneqq
     \kB  %
     \bra{\bx}[\ln p(\mfh), \hat {\bf v} ]_+\ket{\bx}
     /2
       $,
where $\hat {\bf v} \coloneqq (\overset\longleftarrow\nabla-\overset\longrightarrow\nabla)/2m\ci$ denotes a velocity operator; it represents the local current density that constitutes the flow of free energy, $I_\Omega$, we have been after. The overall situation becomes most transparent in equilibrium after expressing 
the expectation value $\bj_s(\bx)$ as a sum over all eigenstates, $\ket{\nu}$, of $\mfh$ with eigenvalues $\epsilon_\nu$: 
\begin{align}
    T \bj_s(\bx)\coloneqq   
     \sum_{\nu} [(\epsilon_\nu-\mu)f(\epsilon_\nu) 
     {\color{black}-}\omega_\nu] {\bf v}_\nu(\bx), 
     \label{e3} 
\end{align}
where ${\bf v}_\nu(\bx) \coloneqq \bra{\bx}\ket{\nu} \hat {\bf v} \bra{\nu}\ket{\bf x}$.  
The first two terms resemble the familiar energy current, $\bj_e(\bx)$, 
and $\mu \bj_n(\bx)$; 
the third term is the novel one; it represents the free energy per eigenstate $\omega_\nu \coloneqq \kB T \ln p(\epsilon_\nu)$ weighted with the particle current density associated with this eigenstate at position ${\bx}$.
The correct quantum results shown in Fig.\ \ref{fig_entropychange_comp} (red curve) and Figs.\ \ref{f1}B and \ref{f1}C (red curve) are calculated based on Eq.\ \eqref{e3}.

\subsection*{Recovering the 3rd Law and thermal noise} 
We offer a brief discussion of our results that further clarifies their broad impact. 
We first point out that Eq.\ \eqref{e3} has an equivalent rewriting as 
\begin{align}
     \bj_s(\bx)\coloneqq
     \sum_\nu s(\epsilon_\nu)
     {\bf v}_\nu(\bx), 
     \label{e4a}
\end{align}
where $s(\epsilon){=}-\kB [f(\epsilon)\ln f(\epsilon) {+} p(\epsilon) \ln p(\epsilon)]$ is the fermionic entropy function \cite{Landau_stat_mech}.
This implies that the current density is given by the entropy carried per state weighted with the state's velocity. 
In this formulation, it is obvious that $\bj_s$ vanishes in the limit of low temperatures, as required by the 3rd Law, because the entropy per state vanishes with $T$. 

To relate the heat current to thermal noise, one considers the change of the entropy current with the chemical potential, %
\begin{align}
    T \partial_\mu \bj_s(\bx)= 
     \sum_{\nu} \beta (\epsilon_\nu-\mu)f(\epsilon_\nu)p(\epsilon_\nu)
    {\bf v}_\nu(\bx),
    \label{e4b} 
\end{align}
which is revealing because the factor $f(\epsilon_\nu)p(\epsilon_\nu)$ is well known to describe the equilibrium (charged) particle current noise in conducting wires \cite{Blanter2000}.
The particle noise weighted with $(\epsilon_\nu-\mu)$ describes the energy fluctuations associated with the fluctuations of the occupation numbers, which is nothing but the thermal noise per state. Hence, Eq.\ \eqref{e4b} implies that the growth of the heat current when increasing the chemical potential reflects the extra thermal noise acquired with this increase. 
The formula analogous to Eq.\ \eqref{e4b} derived within the conventional framework exhibits an extra term. This term is obviously unphysical because it remains finite even at zero temperature (see Sec.\ \ref{app:sanity} in \cite{methods}).

\section*{Unitary (microscopic) to dissipative (macroscopic) crossover}

In order to understand how the microscopic formula for entropy flow crosses over \cite{landauerElectricalTransportOpen1987,jarzynskiEqualitiesInequalitiesIrreversibility2011}
into the well known macroscopic formula, 
we analyze the production of entropy in an exactly solvable model.
Specifically, we consider a quantum wire
wherein the conserved flow of entropy under unitary quantum evolution is taken into account using the exact formula for the entropy current of a system of independent quantum particles. In this exact microscopic description of the quantum dynamics, the entropy production due to Joule heating---the key manifestation of the 2nd Law of Thermodynamics in electrical transport---does not arise without the explicit inclusion of thermalization processes.

\begin{figure}
 \centering
\includegraphics[width=1\linewidth]{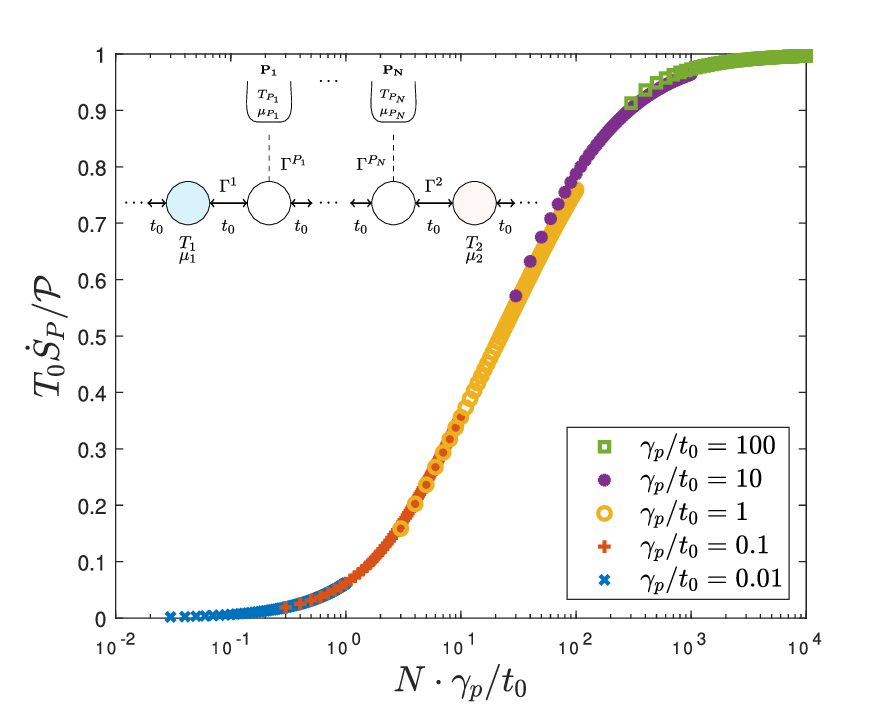}
\caption{{\bf Crossover from unitary to dissipative entropy flow in an exactly solvable model of a quantum wire under electric bias.} The ratio of entropy produced by continuous measurements of $N$ floating thermoelectric probes to entropy due to Joule heating in an infinite tight-binding chain under electric bias with $N \in \{3,\dots ,100\}$ and multiple fixed values of the probe coupling $\gamma_p$. Here $t_0$ is the tight-binding matrix element and
 $T_0$ denotes the absolute temperature of the source and drain reservoirs.
}
\label{fig:sgen1}
\end{figure}

To model this, we equip the quantum wire with a series of floating thermoelectric probes \cite{buttiker89,buttiker88,buttiker2007,Bergfield2014,Shastry2016} along its length (see Fig.\ \ref{fig:sgen1}, inset). In short, these probes have been designed to mimic dephasing processes; they inject entropy into the system as a result of the information obtained via their continuous measurements. 
The corresponding entropy production is described by $\dot S_P=-\sum_n I^{S}_{P_n}$, where $I^S_{P_n}$ denotes the total entropy current into the $n$th probe, which can be obtained, e.g., evaluating Eq.\ \eqref{e4a} within the Landauer-B\"uttiker framework
(see Sec.\ \ref{app:entropy_gen} in \cite{methods}). 
A crossover to dissipative dynamics is expected when the rate of dissipation is comparable to the spacing between discrete energy levels, as the measured  
quantum system would exhibit them, if it were in isolation. 

The crossover between the microscopic and macroscopic descriptions is depicted in Fig.\ \ref{fig:sgen1}, which shows the entropy production associated with a current flow for an increasing number of thermalizing probes.  
As expected, the total entropy production in the wire approaches the macroscopic value associated with Joule heating $\mathcal{P}=IV$ when the dynamics becomes dissipative. At the crossover, the dissipation rate associated with the coupling $\gamma_p$ of the probes to the system is seen to be of the order of the single-particle level spacing $\propto t_0/N$, again confirming our expectations.

\section*{Implications for quantum technology}

A quantum formula for the flow of heat and entropy has been rigorously derived that includes an important quantum term corresponding to a coherent flow of free energy that has been missing since the advent of quantum mechanics a century ago.  This free-energy-flow term is divergence-free for steady-state flows, and thus prior derivations \cite{mahanManyParticlePhysics2000,Imry1,bergfieldThermoelectricSignaturesCoherent2009c} based on continuity equations are agnostic with respect to it.  Only with this quantum term does the entropy current satisfy the fundamental condition that entropy currents must vanish when the entropy itself vanishes.  Importantly, the corrected values for heat dissipated and entropy generated during quantum processes are therefore orders of magnitude smaller than would have been expected based on the conventional formul{\ae}. Our results are not only of fundamental importance, but also of great practical relevance for future quantum technologies. We predict that the thermodynamic limits on the efficiency of quantum machines are far less severe than heretofore %
anticipated.

\clearpage %

\bibliography{main} %
\bibliographystyle{sciencemag}

\section*{Acknowledgments}
We thank Michael Galperin (Tel Aviv University) for useful and inspiring discussions over many years at the Telluride Science Research Center. 
\paragraph*{Funding:}
F.E. has been supported by the Deutsche Forschungsgemeinschaft (DFG, German Research Foundation) 
through research grant SFB 1277, project ID 314695032. 
\paragraph*{Author contributions:}
C.A.S.\ and F.E.\ conceived and supervised the study. M.J., P.K., C.A.S., Y.X., 
and F.E.\ developed the microscopic theoretical model. M.J.\ and P.K.\ produced the data, which was analyzed by M.J., P.K., C.A.S.\, and F.E.. 
M.J., P.K., C.A.S., and F.E.\ wrote the manuscript with contributions from all authors.
\paragraph*{Competing interests:}
There are no competing interests to declare.
\paragraph*{Data and materials availability:}  The datasets and codes used to
 produce the results of this paper will be made available at Zenodo prior to publication.

\subsection*{Supplementary materials}
Materials and Methods\\
Figs.\ S1 to S4\\
References \textit{(39-\arabic{enumiv})}\\ %

\newpage

\renewcommand{\thefigure}{S\arabic{figure}}
\renewcommand{\thetable}{S\arabic{table}}
\renewcommand{\theequation}{S\arabic{equation}}
\renewcommand{\thepage}{S\arabic{page}}
\setcounter{figure}{0}
\setcounter{table}{0}
\setcounter{equation}{0}
\setcounter{page}{1} %

\begin{center}
\section*{Supplementary Materials for\\ \scititle}

    Marco A.\ Jimenez-Valencia, %
    Parth Kumar$^{\ast}$, 
    Yiheng Xu, \\
    Ferdinand Evers, 
    Charles A.\ Stafford\\  
	\small$^\ast$Corresponding author. Email: parthk@arizona.edu%

\end{center}

\subsubsection*{This PDF file includes:}
Materials and Methods\\
Figs.\ S1 to S4\\
References \textit{(39-\arabic{enumiv})}\\

\newpage

\section*{Materials and Methods}

\section{Nonlocal quantum work}
\label{app:nonlocal_work}

\begin{figure}
    \centering
    \resizebox{12cm}{2cm}{%
    \begin{tikzpicture}[scale=0.9
]

			\foreach \x in {-3,0,...,12}
		{	%
  \draw[ultra thick, fill=green!20] (-3,0) circle (1cm);
			\draw[ultra thick, fill=red!20] (\x,0) circle (1cm);
			\draw[ultra thick] (\x + 1,0)--(\x + 2,0);

		}
	\node at (-3,0) {\Large $\varepsilon_s(t)$}; %
    \node at (-1.5,-.5) { $V(t)$};
    \node at (1.5,.5) {\Large $t_0$};
    \node at (4.5,.5) {\Large $t_0$};
     \node at (7.5,.5) {\Large $t_0$};
      \node at (10.5,.5) {\Large $t_0$};
	\node at (0,0) {\Large $\varepsilon_0$};
	\node at (3,0) {\Large $\varepsilon_0$};
	\node at (6,0) {\Large $\varepsilon_0$};
	\node at (9,0) {\Large $\varepsilon_0$};
	\node at (12,0) {\Large $\varepsilon_0$};\node at (13,-1.5) {\Large $\infty$};\node at (3.3,-1.5){\Large $[T,\mu]$};
	\draw [very thick, -latex] (11.6,-1.5)--(12.4,-1.5);
	
\end{tikzpicture}
}
    \caption{{\bf Schematic diagram of a resonant-level $\varepsilon_s(t)$ strongly coupled to a reservoir.} The reservoir is modeled as a semi-infinite tight-binding chain [Eq.\ \eqref{eq_oqs_ham_gen}].  Throughout the driving protocols, the reservoir is held at fixed temperature $T$ and chemical potential $\mu$, and the on-site energy for the chain is $\varepsilon_0=0$. 
    }
    \label{fig_schm_rlm-tbc}
\end{figure}
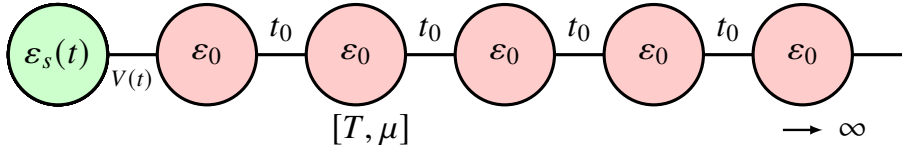

The Hamiltonian of the open quantum system is 
\begin{equation}\label{eq_oqs_ham_gen}
    H(t)=H_S(t)+H_R+H_{S\mbox{-}R}(t)\,,
\end{equation}
where $ %
    H_{S}(t)=\varepsilon_{s}(t)d^{\dagger}d$,
$ %
    H_{R}= %
    t_0\sum_{j=1}^\infty (c_{j}^{\dagger}c_{j+1}+\mbox{h.c.})
$ %
models the reservoir as a semi-infinite tight-binding chain with hopping integral $t_0$,
and the interface between system and reservoir is modelled by 
$ %
H_{S\mbox{-}R}(t)=V(t)d^{\dagger}c_{1}+\mbox{h.c.}$
(see Fig.\ \ref{fig_schm_rlm-tbc} for a schematic of the system).

\begin{figure}
    \centering
    \includegraphics[width=0.9\linewidth]{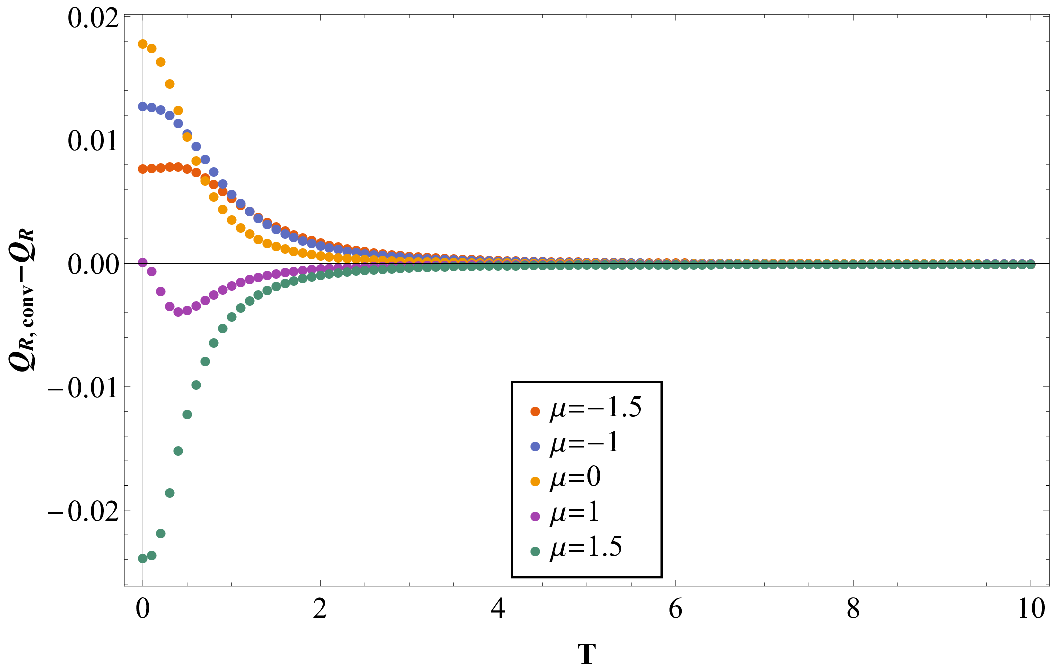}

    \caption{{\bf Difference in conventional and correct formul\ae\ for heat.} The discrepancy in heat dissipated in the reservoir as a function of temperature for several values of the chemical potential $\mu$. Here the level is driven $\varepsilon_s$: $1 \rightarrow 1.5$, and the coupling is fixed at $V=1$. The hopping element $t_0=1.25$.}
    \label{fig_heat_diff}
\end{figure}

The correct quantum formula for the heat generated in the reservoir is $Q_R = Q_{R,{\rm conv}}-\Delta \Omega_R$,
where $\Delta\Omega_R$ is the change in reservoir grand potential due to nonlocal quantum work and $Q_{R,{\rm conv}}=\Delta E_R-\mu \Delta N_R$ is the conventional heat formula.
The rate of nonlocal work done on the reservoir is \cite{kumarWorkSumRule2024a}
\begin{eqnarray}\label{eqn_nonlocal_qwork_sys}
    \delta\dot{\Omega}^{(1)}_R = \frac{1}{2}\langle \dot{H}_{S\mbox{-}R}(t)\rangle &-&\int \frac{d\epsilon}{\pi}\, \mathbb{ImTr}\Bigg\{\frac{\partial \mathcal{G}^{A(0)}(t,\epsilon)}{\partial t}\frac{\partial \Sigma^{A}(t,\epsilon)}{\partial\epsilon}\nonumber \\ &-&\frac{\partial \mathcal{G}^{A(0)}(t,\epsilon)}{\partial\epsilon}\frac{\partial \Sigma^{A}(t,\epsilon)}{\partial t}\Bigg\}\omega(\epsilon)\,,
\end{eqnarray}
where $\mathcal{G}^{A(0)}(t,\epsilon)$ and $\Sigma^{A}(t,\epsilon)$ are the advanced Green's function and self-energy of the system, respectively, evaluated in the quasi-static driving limit, and $\omega(\epsilon)=-\frac{1}{\beta}\ln(1+e^{-\beta(\epsilon-\mu)})$ is the contribution to the grand potential from an orbital of energy $\epsilon$.
The second term on the RHS of Eq.\ \ref{eqn_nonlocal_qwork_sys} may be interpreted as an instantaneous flow of free energy into the reservoir induced by the time-dependent external drive, while the nonlocality of the first term on the RHS is trivial since $H_{S\mbox{-}R}$ is itself nonlocal.  In Figs.\ \ref{fig_entropychange_comp} and \ref{fig_heat_diff}, we set $\dot{V}(t)=0$ so that this trivial nonlocality is absent.

Fig.\ \ref{fig_heat_diff} plots the difference between the conventional and correct heat 
formul{\ae} as a function of temperature $T$ of the reservoir. Clearly the difference is significant for low temperatures $T\lesssim V$, indicating the setting in of a quantum phenomenon at low temperatures---nonlocal quantum work. Finally, Fig.\ \ref{fig_entropychange_comp} compares the conventional and correct entropy changes in the reservoir during the drive as a function of temperature. The plot clearly shows that the conventional formula (blue dots) gives a divergent change in entropy as $T\rightarrow0$ in stark violation of the 3rd Law of Thermodynamics, whereas the the correct formula (red triangles) gives $\Delta S\rightarrow0$ as $T\rightarrow0$, in accordance with the 3rd Law. 

\section{Green's function method for current densities}
\label{app:SGF}

The particle current density %
may be calculated using nonequilibrium Green's functions \cite{dattaElectronicTransportMesoscopic1995,stefanucciNonequilibriumManyBodyTheory2013} as
	\begin{equation}\label{eq:jn}
		\bj_n(\bx)=\frac{\hbar}{2m}\lim_{\bx'\rightarrow\bx}\int \frac{d\omega}{2\pi}
        [
        \nabla'-\nabla +\frac{2iq}{\hbar}\bA(\bx)
        ]
        G^{<}(\bx,\bx';\omega), 
	\end{equation}
where $\bA$ is the magnetic vector potential, $q$ is the electron charge,
$G^<=G^R \Sigma^< G^A$, $G^R$ and $G^A$ are the retarded and advanced Green's functions, respectively, and $\Sigma^< = i\sum_\alpha \Gamma^\alpha f_\alpha(\omega)$ is a source term of particles injected from the reservoirs, with $\Gamma^\alpha$ and $f_\alpha(\epsilon)=\{\exp[(\epsilon-\mu_\alpha)/k_B T_\alpha]+1\}^{-1}$ the tunneling-width matrix and Fermi-Dirac distribution for reservoir $\alpha$, respectively.  The conventional entropy current density shown in 
Fig.\ \ref{f1}A was calculated from a formula related to Eq.\ \eqref{eq:jn} by inclusion of an additional factor of $(\omega -\mu)$ in the integrand.

The entropy current density for a system of independent quantum particles shown in Fig.\ \ref{f1}B was calculated in an analogous fashion as
\begin{equation}
		\bj_s(\bx)=\frac{\hbar}{2m}\lim_{\bx'\rightarrow\bx}\int \frac{d\omega}{2\pi}
        [
        \nabla'-\nabla +\frac{2iq}{\hbar}\bA(\bx)
        ]
        G^{S}(\bx,\bx';\omega),
	\end{equation}
where $G^S\equiv G^R \Sigma^S G^A$ is an entropy Green's function describing the unitary flow of entropy in a quantum scattering problem.  Here $\Sigma^S(\omega) = i\sum_\alpha\Gamma^\alpha s_\alpha(\omega)$ can be interpreted as a source term of entropy injected from the reservoirs, with $s_\alpha(\omega)=-k_B\left[ f_\alpha \log f_\alpha+(1-f_\alpha)\log(1-f_\alpha) \right]$.

\section{Entropy generation}
\label{app:entropy_gen}

\begin{figure}
\includegraphics[width=1\linewidth]{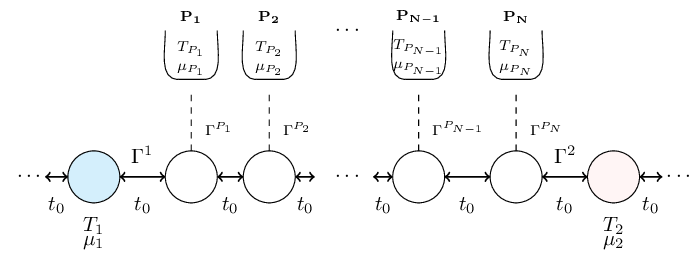}
	\caption{ {\bf Model system consisting of an infinite %
    tight-binding chain coupled to $N$ floating thermoelectric probes.}  The probes serve as sources of decoherence and thermalization.  
    Information acquired through the probes' continuous measurements results in entropy injected into the wire.
    The semi-infinite sections to the left and right of the probe region represent the source and drain reservoirs. \label{fig:model}  }
\end{figure}

As an example to illustrate how our microscopic (unitary) formula for entropy flow leads in the macroscopic limit to dissipation and the 2nd Law, we analyzed the phenomenon of Joule heating in an exactly solvable model of transport in a quantum wire, shown schematically in Fig.\ \ref{fig:model}.  
The Hamiltonian of the quantum wire is 
\begin{equation}
    H = H_S + H_R + H_P + H_{S\mbox{-}RP}, %
\end{equation}
where %
\begin{equation}
    H_S = t_0\sum_{i=1}^N d^\dagger_i d_{i+1} + \mbox{h.c.} \label{sysh}
\end{equation}
describes the central region of the wire, $t_0$ being the tight-binding hopping matrix element between neighboring sites of the wire, the
reservoirs are described by
\begin{equation}
    H_R = \sum_{\alpha = a,b} t_0 \left( \sum_{j=1}^\infty c_{j,\alpha}^\dagger c_{j+1,\alpha} +\mbox{h.c}. \right), \label{resh} 
\end{equation}
where $\alpha$ labels the reservoirs, $a$ for the source on the left and $b$ for the drain on the right. The labeling is such that the subscript $(1,a)$ corresponds to the site at the interface between the left reservoir and the first site in the system, and $(1,b)$ to the site at the interface between the right reservoir and the $N$th site in the system counted from the first probe. The floating probes are also reservoirs described as metallic Fermi gases
\begin{equation}
    H_{P}=\sum_{\beta=1}^N H_{P_\beta}=\sum_{\beta=1}^{N}\,\sum_{k\in P_\beta} \varepsilon_{k\beta} c_{k,P_\beta}^\dagger c_{k,P_\beta} \label{probeh},
\end{equation}
and the couplings are described by
\begin{equation}
    H_{S\mbox{-}RP} %
    =t_0 \left(c_{1,a}^\dagger d_1 + c_{1,b}^\dagger d_N  \right) + \sum_{\beta=1}^N\sum_{k\in P_\beta} V_{k\beta}c_{k,P_\beta}^\dagger d_\beta +\mbox{h.c.} \label{couplingsh},
\end{equation}
where the first term corresponds to the coupling of the semi-infinite source and drain to the left and right and the second term corresponds to the coupling of the $N$ probes to the corresponding sites in the system.  The coupling matrix of the $n$th probe to the system is taken as $(\Gamma^{P_n})_{ij}=\gamma_p \delta_{in}\delta_{jn}$.
Each of the reservoirs (source, drain, and probes 1 to $N$) is taken to be macroscopic and without coherent backscattering, enforced by an order of limits within scattering theory \cite{barangerElectricalLinearresponseTheory1989,nockelAdiabaticTurnonAsymptotic1993}.

\begin{figure}
 \centering
\includegraphics[width=1\linewidth]{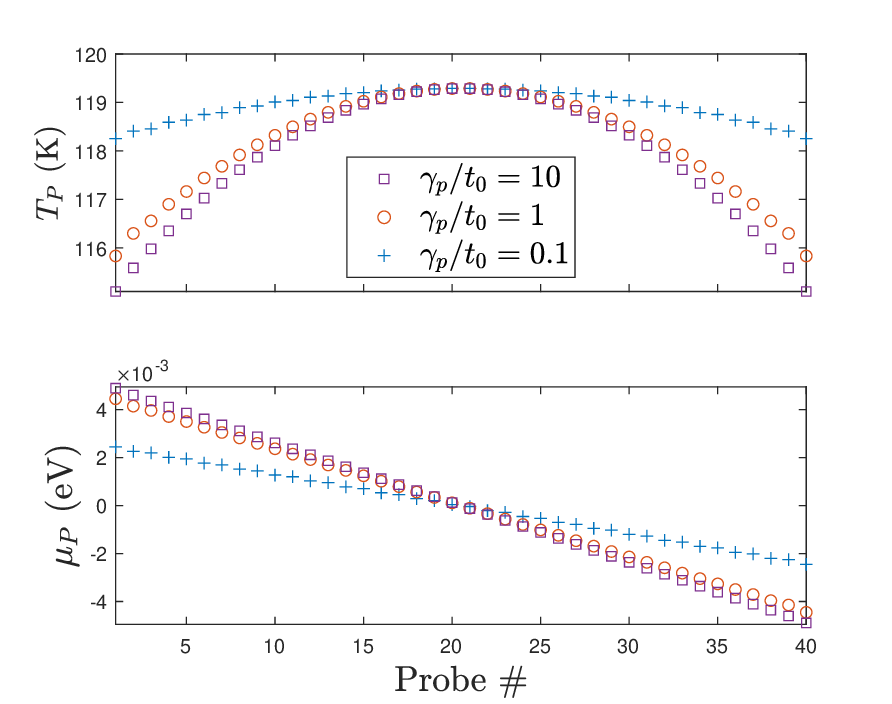}
\caption{ {\bf Temperatures and chemical potentials measured by 40 floating probes in an infinite chain under electric bias.} Results for three values of the probe coupling strength $\gamma_p$ are shown. 
Here source and drain are held at a common temperature $T_0=115$K, the hopping matrix element in the chain is $t_0=2.7$eV, and the electric bias between source and drain is $\Delta \mu=10k_BT_0\approx 0.1$ eV.
}
\label{fig:mts}
\end{figure}

Each floating probe labeled by $P_n$ performs a continuous measurement \cite{buttiker89,buttiker88,buttiker2007,Bergfield2014,Shastry2016} that takes pure states in the wire and returns mixed states, acquiring information about the electronic states in the wire and disposing of that information by injecting entropy into the system, not storing it. These probes serve as neither sources of particles nor energy in the wire, leading to the floating conditions
\cite{buttiker89,buttiker88,buttiker2007,Bergfield2014,Shastry2016} 
\begin{multline}
    I_{P_n}^{(\nu)}(\mu_{P_1},\dots,\mu_{P_N},T_{P_1},\dots,T_{P_N}) = 0\,, \\ 
    \forall\ n=1,\dots, N, \text{ and } \ \nu=0,1, \label{probecond} 
\end{multline}
which are solved \cite{jimenez25d} to determine the probe
chemical potentials and temperatures, 
where the particle ($\nu=0$) and energy ($\nu=1$) currents in Eq.\ \eqref{probecond} are calculated from \cite{buttiker4_1986,Imry1,bergfieldThermoelectricSignaturesCoherent2009c}
\begin{equation}
I^{(\nu)}_\alpha=\frac{1}{h}\int d\epsilon \, \epsilon^\nu\sum_\beta T_{\alpha\beta}\left[f_\beta-f_\alpha\right], \label{BSI}
\end{equation}
where $T_{\alpha\beta}(\epsilon)={\rm Tr}\{\Gamma^\alpha G^R \Gamma^\beta G^A\}$ is the transmission function from reservoir $\beta$ to reservoir $\alpha$.
The temperatures and chemical potentials measured simultaneously by $N=40$ probes for a quantum wire under electric bias, with source and drain reservoirs held at a common temperature $T_0$, are shown in Fig.\ \ref{fig:mts}.

The entropy injected into the quantum wire by the $N$ floating probes is calculated from the following formula for the entropy current flowing into reservoir $\alpha$
\begin{equation}
     I^{S}_{\alpha}=\frac{1}{h}\int d\epsilon \sum_\beta T_{{\alpha}\beta}\left[s_\beta\left(\epsilon\right)-s_\alpha\left( \epsilon\right)\right], \label{unitarys}
\end{equation}
which may be derived from Eq.\ \eqref{e4a} by performing surface integrals over the junctions between the quantum wire and the reservoirs, and using the entropy Green's function introduced in Sec.\ \ref{app:SGF}. %
The total rate of entropy injected by the $N$ probes is
\begin{equation}
    \Dot{S}_P = -\sum_n I_{P_n}^S \label{spsom}.
\end{equation}
This is to be compared with the total (macroscopic) rate of entropy production due to Joule heating
\begin{equation}\label{eq:Joule}
    \frac{{\cal P}}{T_0}= \frac{I_1^{(0)}(\mu_2-\mu_1)}{T_0},
\end{equation}
where ${\cal P}=IV$ is the electrical power supplied by the source and drain electrodes.  The ratio of Eqs.\ \eqref{spsom} and \eqref{eq:Joule} is plotted for a range of $N$ and $\gamma_p$ values in Fig.\ \ref{fig:sgen1}.

\section{Sanity checks} 
\label{app:sanity}

Differentiating the conventional entropy current density, we find
\begin{eqnarray}
    T \partial_\mu \bj_{s,{\rm conv}}(\bx) & = & %
     \sum_{\nu} \beta (\epsilon_\nu-\mu)f(\epsilon_\nu)p(\epsilon_\nu)
    {\bf v}_\nu(\bx) \nonumber \\
    & - & \sum_\nu f(\epsilon_\nu){\bf v}_\nu(\bx).
\end{eqnarray}
The last term is clearly unphysical as it remains finite as $T\rightarrow 0$, in violation of the 3rd Law.

Finally, it is instructive to show that the divergence of $\bj_\Omega$ vanishes in equilibrium. This is already clear by construction, but it also follows directly from \eqref{e3}:
\begin{align}
\text{div}\bj_\Omega(\bx) &=     
    \sum_{\nu} \omega_\nu 
    \bra{\bx}\ket{\nu} (\cev\nabla\hat {\bf v} - \hat {\bf v}\vec\nabla)\bra{\nu}\ket{\bf x}
    \nonumber\\
    &= \frac{1}{\ci}
    \sum_{\nu} \omega_\nu 
    \bra{\bx}\ket{\nu} (-\cev\mfh + \vec\mfh)\bra{\nu}\ket{\bf x}\nonumber\\
        &= -\frac{\kB T}{\ci}\bra{\bx}[\ln p(\mfh),\mfh]\ket{\bx} =0 \label{e4}
\end{align}
In quasi-stationary non-equilibrium, however, the distribution $p$ ceases to commute with $\mfh$, in general. Also for this case we maintain the definition \eqref{e2} and conclude that 
$\bj_\Omega$ can develop a divergence and contribute also to the entropy transport between a system and its environment
(see Fig.\ \ref{fig_entropychange_comp}).

\end{document}